\documentclass[%
 reprint,
 amsmath,amssymb,
 aps,
]{revtex4-2}

\usepackage{graphicx}
\usepackage{dcolumn}
\usepackage{bm}

\usepackage{verbatim}

\usepackage[normalem]{ulem}
\usepackage{color}

\newcommand{\ba}{\mathbf{a}}
\newcommand{\Koop}{\mathcal{K}}
\newcommand{\Kapp}{\mathbf{K}}

\begin{document}

\preprint{APS/123-QED}

\title{Predicting extreme events in a data-driven model of turbulent shear flow using an atlas of charts}

\author{Andrew J. Fox}
\author{C. Ricardo Constante-Amores}
\author{Michael D. Graham}%
 \email{mdgraham@wisc.edu}
\affiliation{Department of Chemical and Biological Engineering,\\
University of Wisconsin-Madison,\\
Madison, WI 53706, USA
}%

\date{\today}

\begin{abstract}

Dynamical systems with extreme events are difficult to capture with data-driven modeling, due to the relative scarcity of data within extreme events compared to the typical dynamics of the system, and the strong dependence of the long-time occurrence of extreme events on short-time conditions.
A recently developed technique [Floryan, D. \& Graham, M. D. Data-driven discovery of intrinsic dynamics. Nat Mach Intell \textbf{4}, 1113–1120 (2022)], here denoted as \emph{Charts and Atlases for Nonlinear Data-Driven Dynamics on Manifolds}, or CANDyMan, overcomes these difficulties by decomposing the time series into separate charts based on data similarity, learning dynamical models on each chart via individual time-mapping neural networks, then stitching the charts together to create a single atlas to yield a global dynamical model. 
We apply CANDyMan to a nine-dimensional model of turbulent shear flow between infinite parallel free-slip walls under a sinusoidal body force  [Moehlis, J., Faisst, H. \& Eckhardt, B. A low-dimensional model for turbulent shear flows. New J Phys \textbf{6}, 56 (2004)], which undergoes extreme events in the form of intermittent quasi-laminarization and long-time full laminarization.
We demonstrate that the CANDyMan method allows the trained dynamical models to more accurately forecast the evolution of the model coefficients than a single-chart model, reducing the error in the predictions as the model evolves forward in time.
The technique exhibits more accurate predictions of extreme eventsthan a single-chart model, capturing the frequency of quasi-laminarization events and predicting the time until full laminarization more accurately than a single neural network.

\end{abstract}

\maketitle

\section{Introduction}

Real world dynamical systems often produce unusual behaviors in the form extreme events.
These extreme events are characterized by a dissimilarity to the typical dynamics of the system, usually greater in scope or scale, that occur relatively infrequently compared to the typical dynamics.
Common examples include rogue waves in the ocean \citep{Dysthe2008}, extreme weather patterns such as hurricanes and tornadoes \citep{Easterling2000,Majda2012}, and intermittency in turbulent flows \citep{Platt1991}.
While extreme events are a consequence of the same dynamical system that governs the non-extreme state, they are often difficult to forecast using data-driven modeling.
The relative scarcity of data within extreme events both limits the overall observations of the extreme events on which to train the model and reduces the relative influence of extreme event behavior on data-driven model training.
Thus, creating a data-driven model that can accurately capture extreme events remains a active challenge.

Recent studies have proposed various techniques for analyzing and forecasting the occurrence of extreme events. 
\citet{Guth2019} developed a probabilistic framework for the use of indicator observables as predictors of the extreme events.
\citet{Ragone2021} supplemented climate model simulations with a rare event algorithm to examine and more accurately capture the increasing frequency of extreme heatwaves in Europe.
\citet{Blanchard2022} built a machine learning framework to correct a biased climate model to produce better forecasts of extreme events.
\citet{Mendez2022} applied probabilistic models toward predicting indirect spreading of wildfires by wind to improve forecasts of new wildfire locations.
\citet{Gome2022} applied a rare-event algorithm to analyze the transition between states in turbulent pressure-driven flow and more efficiently predict passage time between states.
While these studies improved predictions of extreme events, they primarily corrected and supplemented the forecasts of existing models; we will instead aim to develop an improved model.

One attractive test case of a dynamical system with extreme events is the nine-dimensional model for turbulent flow developed by Moehlist, Faisst, and Eckhardt (MFE) \citep{Moehlis2004}.
The MFE model, an extension of a model by Waleffe \citep{Waleffe1997}, governs the evolution of nine amplitudes of combinations of spatial Fourier modes describing an incompressible turbulent shear flow between infinite parallel free-slip walls under a sinusoidal body force.
These nine modes provide a minimal description of the mechanisms for self-sustenance in turbulence, allowing the resulting flow field to display realistic turbulent dynamics. In particular, the model
displays features consistent with turbulence in the transition region, namely long periods of turbulent behavior with infrequent quasi-laminarization events (also called quiescent \citep{Hamilton1995} or hibernating \citep{Xi2010} intervals) and ultimately full laminarization \citep{Hamilton1995, Xi2010, Hof2006}.
These quasi- and complete relaminarizations will be the extreme events considered in the present work, in which we use time series from the MFE model as ``data" with which to develop a data-driven model.

In recent years, several attempts have been made to reproduce the dynamics of the MFE model (and other flow systems) through data-driven techniques based on neural networks (NNs).
Neural networks are a powerful data-driven modeling technique that has been shown to accurately recreate the dynamics of systems such as the viscous Burgers equation\citep{Linot2023}, the Kuramoto-Sivashinksy equation\citep{Linot2020, Linot2022},  and Kolmogorov flow\citep{Perez2022}.
\citet{Srinivasan2019} developed both feedforward neural networks (FNNs) and long short term memory (LSTM) networks to recreate the MFE model as discrete-time maps.
While the FNNs were unable to reproduce the model, LSTMs were able to accurately reconstruct long-time behaviors of the full-field velocity statistics.
This problem was revisited by  \citet{Eivazi2021}, where the reconstruction via a LSTM network was compared to predictions generated via a Koopman-operator-inspired framework with nonlinear forcing. 
In their approach, the observables incorporate time-delay-embeddings, and they impose a nonlinear forcing \citep{khodkar2019koopmanbased,Brunton2017}.
Their work demonstrated that this framework could reproduce short-time and long-time statistics as well or better than the LSTM networks. (We further discuss the Koopman operator approach to dynamics below, for the moment simply noting that the original Koopman operator formalism is linear and Markovian, neither of which property is exhibited by the methodology of \cite{Eivazi2021}.)
 \citet{Pandey2020} introduced the use of reservoir computing in the form of an echo state network (ESN), to reproduce the MFE model as a discrete-time map, and provided comparisons to both a FNN and a LSTM network.
The LSTM network and the ESN were shown to perform similarly, with both adequately capturing the full-field velocity statistics, while again the FNN was shown to perform appreciably worse.
\citet{Racca2022} specifically examined the ability of an ESN to forecast the occurrence of an extreme event within a future time window.
They determined that their data-driven model could accurately forecast extreme event episodes far into the future without incorrectly predicting false quasi-laminarization events. 
\citet{Pershin2021} assessed the ability of an ESN to forecast time until full laminarization.
They showed that their model could adequately reproduce the lifetime distribution of the MFE data, correctly predicting the probability of an arbitrary MFE time series remaining in the turbulent state some time in the future. 
These studies only successfully modeled the MFE equations through the use of non-Markovian models, which forecast the future state through input of the current and past states.
As the MFE model is itself Markovian, we will instead endeavor to model the MFE data with a Markovian dynamical system.

Specifically, we will use a recently developed method that will be denoted here as \textit{Charts and Atlases for Nonlinear Data-Driven Dynamics on Manifolds} (CANDyMan) \citep{Floryan2022,FloryanRepo}.
CANDyMan operates by decomposing the data distribution in state space into separate regions called charts with a clustering algorithm, learning local dynamical models in each chart using FNNs, then stitching together the charts to create a single atlas containing the global dynamical model. This approach is quite distinct from Cluster-Based Network Modeling \citep{Fernex2021}, where clustering is used to construct Markov chains modeling transitions between cluster centroids.  Here we are constructing deterministic dynamical systems.
This technique has been previously applied to dimension reduction problems, accurately learning reduced order dynamical models whose dimension is equal to the intrinsic dimensionality of the system  \citep{Floryan2022}.
The use of multiple charts allows low-dimensional manifolds embedded in high dimensional space to be broken down into locally low dimensional structures, capturing the dynamics of a system with the minimal number of dimensions, in a way that single chart methods cannot.
Here, we do not perform dimension reduction, but rather utilize the clustering of data to break down the dynamical system into separate regions representing extreme and non-extreme states.
By learning the dynamics in the extreme region separately and independently from the non-extreme regions, CANDyMan inherently overcomes the imbalance of extreme vs non-extreme information and thus the limited influence of extreme events in data driven model training. 

Here, we will use CANDyMan to reconstruct the dynamics of the MFE model.
A data set containing time series of the MFE amplitudes will be decomposed using $k$-means clustering into atlases containing between one and five charts. 
We will train deep neural networks to reconstruct the time evolution of the MFE amplitudes within each of the charts, then stitch them together to create five global models.
To assess the accuracy of the models, we will first consider their ability to reconstruct the turbulent flow field.
Next, we will analyze their performance in reproducing short-time and long-time statistics,  and  we will compare our framework  with a   Koopman framework in which the observable evolve under linear dynamics. Finally, we will assess the extreme event forecasting of the data-driven models by determining the statistical accuracy of forecasting extreme event occurrences and comparing predicted laminarization lifetime distribution to the true data. 


\section{Formulation}

The MFE model is a severely truncated Fourier Galerkin approximation to the Navier-Stokes equations (NSE) for incompressible flow between two free-slip walls and driven by a spatially sinusoidal body force.
The flow is composed of nine combinations of spatial Fourier modes $\mathbf{u}_i (\mathbf{x})$, describing the basic profile, streaks, and vortices, as well as interactions between them. 
The velocity field at position $\mathbf{x}$ and time $t$ is given by a superposition of the nine modes as $\mathbf{u} (\mathbf{x},t) = \sum\limits^9_{i=1} a_i(t) \mathbf{u}_i (\mathbf{x})$.
The mode amplitudes $a_i(t)$ satisfy a system of nine ordinary differential equations (ODEs), generated through Galerkin projection, whose explicit form is given in \citet{Moehlis2004}. 
Our study considers a domain of size $L_x \times L_y \times L_z$, with infinite, parallel walls at $y = -L_y/2$ and $y = L_y/2$ and periodic boundaries $x=0$, $x=L_x$, $z=0$, and $z=L_z$; $x$, $y$, and $z$ are the streamwise, wall-normal, and spanwise coordinates, respectively.
The domain size of $L_x=4\pi$, $L_y=2, Lz=2\pi$ was used, with a channel Reynolds number of 400; these parameters produce turbulent behavior of suitable length for data-driven model development \citep{Srinivasan2019}.

\begin{figure}
\centering
\includegraphics[width=0.45\textwidth]{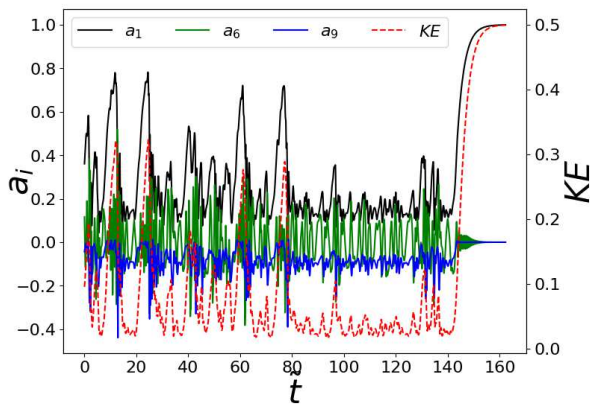}
\caption{Evolution of three amplitudes, $a_1, a_6, a_9$ and corresponding kinetic energy from one time series of the MFE data set.}
\label{MFE_Example}
\end{figure}

As training data, we generated 100 unique time series from a fourth-order Runge-Kutta integration of the MFE equation, with a time step of $0.5$.
Each time series encompasses the transient turbulent state, consisting of turbulent intervals interspersed with quasi-laminarization events, with terminal laminarization occurring at long time.
We will often characterize the flow using the \emph{total} kinetic energy (KE), given by $KE=\frac{1}{2} \sum^9_{i=1} a_i^2$.
Therefore, the turbulent state is \textit{low} energy while the laminar is \textit{high} energy.
Every time series collapses to the known laminar fixed point $a_i=\delta_{i1}$. To generate the time series, initial conditions of eight of the amplitudes were given as follows: $(a_1, a_2 ,a_3 ,a_5 ,a_6 ,a_7 ,a_8 ,a_9) = (1, 0.07066, -0.07076, 0, 0, 0, 0, 0)$.
The initial value of $a_4$ was arbitrarily generated in the range $[-0.1, 0.1]$.
These initial conditions were previously demonstrated to generate chaotic dynamical data with quasi-laminarization events \cite{Srinivasan2019}.
We will report all results in units $\tilde{t} = t/\tau_L$, where $\tau_L$ is the  Lyapunov time for the system; in the original nondimensionalization $\tau_L \approx 61$ \citep{Racca2022}. 
The first 1000 time steps of each time series were discarded to eliminate dependence on initial conditions.
Each time series was evolved until the laminar state was reached, taking a varying number of time steps depending on the initial conditions; the resulting training data set consisted of approximately 2 million snapshots of MFE amplitudes and has a mean lifetime of $164 \tau_L$.
Amplitudes and $KE$ from a randomly chosen time series are shown in Fig.~\ref{MFE_Example}.

In this study, we examined the behavior of multi-chart models with between two and five charts, as well as a standard approach with one global model -- the ``one-chart" limit of CANDyMan.
Here, the one chart refers to a global state space representation by a single dynamical model, whereas in the multi-chart models the separate charts create independent local representations of the state space which are then combined to form a single global model.
The dynamical system data is first clustered into $k$ charts via $k$-means clustering, which partitions a data set into $k$ clusters, minimizing the within-cluster variance \citep{MacQueen1967, Forgy1965}.
Other clustering techniques, such as $k$-nearest neighbors \citep{Fix1989} or single-linkage clustering \citep{Everitt2011}, could be used, provided the clustering technique produces charts that encompass contiguous regions of the state space.  Furthermore, machine learning techniques involving clustering have recently appeared, including Chart Auto-Encoders \citep{Schonscheck2020}, where clustering of the data is learned simultaneously with local dimension reduction, and Mixture Models of Variational Autoencoders \citep{Alberti2023}, where clustering is performed with mixture models.
We selected $k$-means clustering for this study due to its simplicity for implementation and its previous success in modeling dynamical systems with the CANDyMan method \citep{Floryan2022}.
The clusters are then augmented so that they overlap, by locating the $k_{NN}$ nearest neighbors to each data point in a cluster by Euclidean distance and adding these to the original cluster.
This creates overlap regions between neighboring clusters, providing transition regions in which the dynamics are described in multiple charts and allowing for the movement into and out of the region to be handled by the separate local models.

Then, in each augmented chart, we generated discrete-time models of the form $a^{(j)}(t+\tau)=F^{(j)}(a^{(j)}(t);\theta^{(j)})$, 
where $a^{(j)}(t)\in\mathbb{R}^9$ is the representation of the state in chart $j$, the discrete time step is $\tau = 0.5$,  and $F^{(j)}$ is the corresponding discrete-time map, which takes the form of a FNN.
The quantities $\theta^{(j)}$ are the neural network weights for $F^{(j)}$, which are learned from the data using a standard stochastic gradient descent method and trained to minimize the loss function $L^{(j)} = \langle ||a^{(j)}(t) - \tilde{a}^{(j)}(t)||_2 \rangle$, where  $\langle \cdot \rangle$ is the average over the training data.
To ensure that the comparison between different numbers of charts was standardized, each global model contains the same number of total neurons, $N_T = 1800$; a system of $k$ charts would then use $N_N = N_T/k$ neurons in each local model, each containing four fully-connected hidden layers of $N_N/6$, $N_N/3$, $N_N/3$, and $N_N/6$ of the total number of local neurons, respectively.
Increasing the total number of neurons beyond $N_T = 1800$ was not found to significantly impact the performance of the data-driven models.
Each neural network was trained using a learning rate scheduler with an initial learning rate of $0.01$, decaying at a rate of $0.9$ every $2000$ steps.
Each model was then trained for 100 epochs, which was found to accurately reproduce the training data while avoiding overfitting.
The computational cost of training global models containing multiple local models is no greater than that of training a single global model, as the total trainable parameters, total amount of training data, and training procedure is held constant between the different models; additionally, the multi-chart models could potentially be trained in less real time by training models in each chart in parallel.

Finally, we briefly describe the methodology used for the Koopman predictions. For a Markovian autonomous deterministic dynamical system with state $\ba(t)$, the Koopman operator $\Koop_{\delta t}$ describes the evolution of an arbitrary observable $G(\ba)$ from time $t$ to time $t+\delta t$: $G(\ba(t+\delta t)=\Koop_{\delta t}G(\ba(t))$ \cite{koopman1931,Lasota.1994.10.1007/978-1-4612-4286-4}. The Koopman operator is linear and time-independent, so evolution of the observables  of the state can be expressed as a sum (or integral, if $\Koop_{\delta t}$ has a continuous spectrum) of ``Koopman modes" with complex-exponential time dependence. The tradeoff for gaining linearity is that $\Koop_{\delta t}$ is also infinite-dimensional, requiring for implementation some finite-dimensional truncation of the space of observables. Here we use the ``extended dynamic mode decomposition-dictionary learning'' (EDMD-DL) approach \citep{Qianxiao}.
Given a vector of observables $\boldsymbol{\Psi}(\ba(t))$, now there is an approximate matrix-valued Koopman operator $\Kapp$ such that the evolution of observables is approximated by $\boldsymbol{\Psi}(\ba(t+\delta t))=\Kapp\boldsymbol{\Psi}(\ba(t))$. The EDMD-DL approach aims to simultaneously learn the operator $\Kapp$ and the best set of observables $\boldsymbol{\Psi}(\ba(t))$, represented as neural networks, to accurately approximate the evolution of the system.
The key idea behind finding $\Kapp$ is to determine the  linear operator which best maps between corresponding  pairs of observables (in a least-squares sense).
Given a matrix of observables, whose columns are the vector of observables at different times,
$\boldsymbol{\psi}(t)=\left [ \boldsymbol{\Psi}(\ba(t_1))~~ \boldsymbol{\Psi}(\ba(t_2)) ~~\cdot \cdot \cdot \right ]$ and its corresponding matrix at  $ t + \delta t$,
$\boldsymbol{\psi}(t+\delta t)=\left [ \boldsymbol{\Psi}(\ba(t_1 +\delta t))~~ \boldsymbol{\Psi}(\ba(t_2 +\delta t)) ~~\cdot \cdot \cdot \right ]$, 
the approximate matrix-valued Koopman operator is defined as 
$\Kapp= \boldsymbol{\psi}({t + \delta t }) \boldsymbol{\psi}(t)^+ $, 
where $+$ superscript denotes the pseudo (Moore-Penrose) inverse.
Our method relies on automatic differentiation to find $\Kapp$ and the set of observables simultaneously. 
In this study, we create a set of observables  of 100 elements in addition to the  state. 
We select 3 hidden layers with 100 neurons, each hidden layer is followed by an activation function with  ELU.
We have thoroughly varied the numbers of observables, and the neural network architecture finding no difference in the short- and long-time tracking of the results.

\section{Results and Discussion}

\subsection{Distribution of data into clusters}

\begin{figure}
\centering
\includegraphics[width=0.45\textwidth]{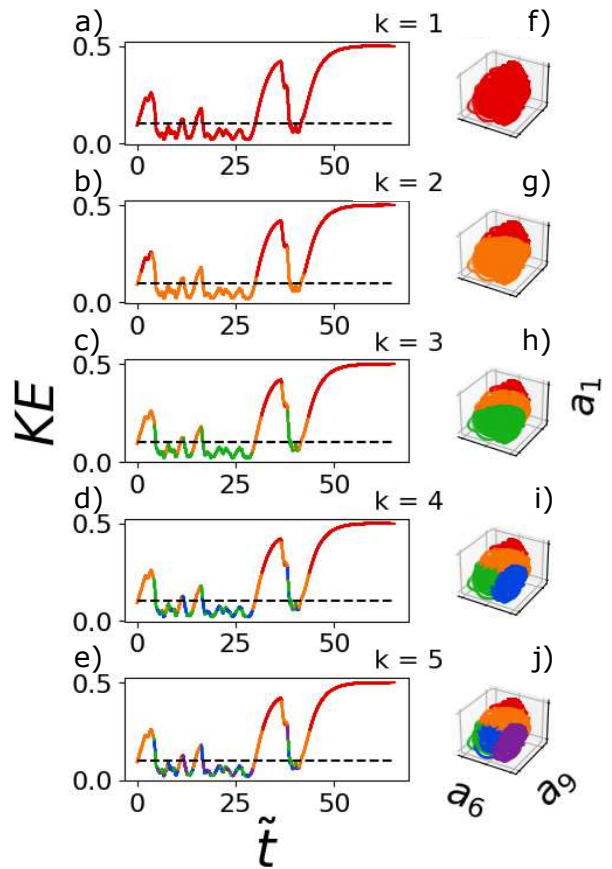}
\caption{Clustering of a randomly selected trajectory of kinetic energy (a-e) and the projection of the clustering of the first, sixth, and ninth MFE amplitude (f-k) for one to five charts, color coded by cluster.}
\label{Clustering}
\end{figure}

Insight into the number of charts necessary for properly reconstructing the MFE data can be gained by observing the clustering of the training data set.
Fig.~\ref{Clustering} shows how one trajectory from the data set is partitioned when we use different numbers of charts, in terms of (a-e) the time series of $KE$ and (f-j) state space projections onto amplitudes $a_1, a_6, a_9$.
With two charts, the  data partitions into one cluster covering the low-energy (turbulent) non-extreme states and the second containing the high-energy extreme (quasi-laminar, laminarizing) states.
When three charts were used, the clusters are further segmented, with one covering the low-energy turbulent state, the second primarily consisting of the transition into quasi-laminarization and laminarization events, and the third consisting mainly of the high energy components of these events.
Clustering into four charts breaks down the low-energy region into two separate clusters that maintain relatively distinct. 
When the data was clustered into five charts, the distinction between the charts in the low-energy turbulent regime decreased and the charts containing the turbulent states were described by increasingly similar centroids.

\subsection{Trajectory predictions and time-averaged statistics}

The performance of the data-driven models was evaluated on their ability to reconstruct the evolution of the MFE model amplitudes.
Two test data sets were generated for comparison between the MFE dynamics and the single- and multi-chart data-driven models.
or trajectory predictions, 1000 trajectories of MFE amplitudes were generated from arbitrary initial conditions and time-integrated for 10 Lyapunov times, with the same initial conditions separately evolved forward using the generated data-driven models for the same length of time; this will henceforth be denoted as data set A.
The purpose of this data set is to determine the short-time precision of the predictions generated by the single- and multi-chart models, regardless of any observed or predicted laminarization.

For time-averaged statistics, 100 trajectories of MFE amplitudes were generated from random initial conditions and time-integrated for 100 Lyapunov times or until a laminarization event occurred, with the initial conditions separately evolved forward using the generated data-driven models and the same ending criteria; this will henceforth be denoted as data set B.
The purpose of this data set is to assess the accuracy of the predicted long-time turbulent state statistics, and as such removes any observed or predicted laminarization.
As with the creation of the training data, the first 1000 time steps of each time series following initialization were discarded for both test data sets.

\begin{figure}
\centering
\includegraphics[width=0.45\textwidth]{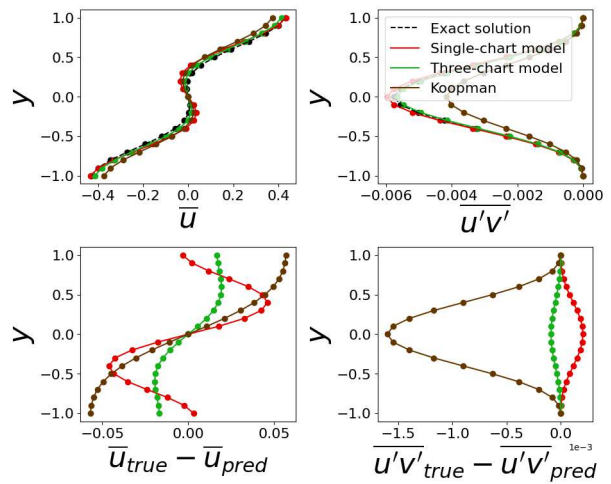}
\caption{(a) Mean streamwise velocity  and (b) Reynolds shear stress  of the full field of the testing data and of the reconstruction of the MFE model by the single- and three-chart model and by the Koopman model, with (c-d) the difference between the true values and the predicted values.}
\label{Flow_Statistics}
\end{figure}

The data-driven models are first evaluated on their ability to reconstruct the velocity statistics of the turbulent regime, the essential function of the MFE model.
Using data set B, we project the amplitudes on to the spatial Fourier modes of the MFE model and compare the accuracy of the predicted velocity statistics in the turbulent state to the exact solution.
The mean streamwise velocity and Reynolds shear stress were calculated for each data set, as shown in Fig.~\ref{Flow_Statistics}.
The mean streamwise velocity and Reynolds shear stress profiles are practically identical to those created by the turbulent portion of the training data.
As the figure shows, the single-chart model and the Koopman operator capture well the form of the velocity statistics, but fail to accurately capture the exact values.
The three-chart model creates much better predictions, quantitatively capturing the flow profile.

\begin{figure}
\centering
\includegraphics[width=0.45\textwidth]{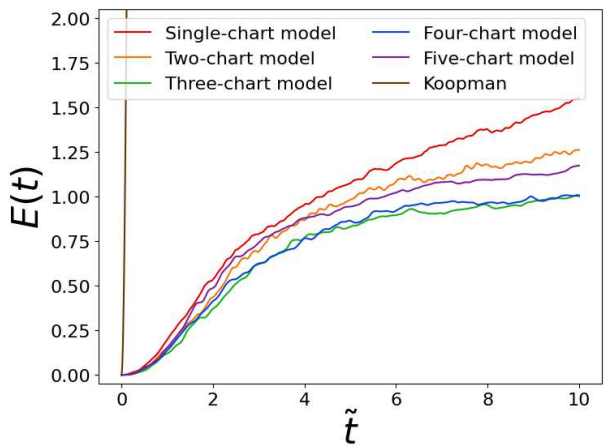}
\caption{Ensemble averaged short-time error tracking of the reconstruction of the MFE model by the single- and multi-chart models, as well as the reconstruction by the Koopman model.}
\label{Ensemble_Averaged_Error}
\end{figure}

Now we turn to the prediction of trajectories. To quantify the performance of the trajectory predictions, we analyzed the data-driven models' ability accurately forecast the evolution of MFE amplitudes. Using data set A, the error in the predictions, $E(t)$, was then calculated for each time series, averaged, and normalized, such that $E(t) = \frac{||a(t) - \tilde{a}(t)||_2}{D}$.
Here, $D$ is the average $L_2$-norm between randomly chosen time instants in the turbulent state.
Fig.~\ref{Ensemble_Averaged_Error} shows $E(t)$ for the single- and multi-chart models and for the Koopman model as a function of time.
All FNN models create accurate predictions for $\sim 0.5 \tau_L$, with the error remaining close to 0, while the error in the predictions by the Koopman model increases rapidly even at short times. The rapid growth in error is caused by a small number of predictions that evolve rapidly away from the true solution, whose large deviations from the true values dominate the error calculation.  The error in the predictions of the single-chart model also grows much more rapidly than the multi-chart models, indicating that the forecasting ability is much stronger in the multi-chart models. Furthermore, the performance of the multi-chart models improves as the number of charts increases from two to three, plateaus from three and four, and diminishes from four to five. This indicates that the three charts are sufficient to improve reconstruction via the CANDyMan technique, while further increases in the number of charts can in fact impair performance. Possible reasons for this are discussed below. As such, for the remainder of the paper, we will focus our comparison between the single- and three-chart models.

\subsection{Prediction of extreme events}

\begin{figure}
\centering
\includegraphics[width=0.45\textwidth]{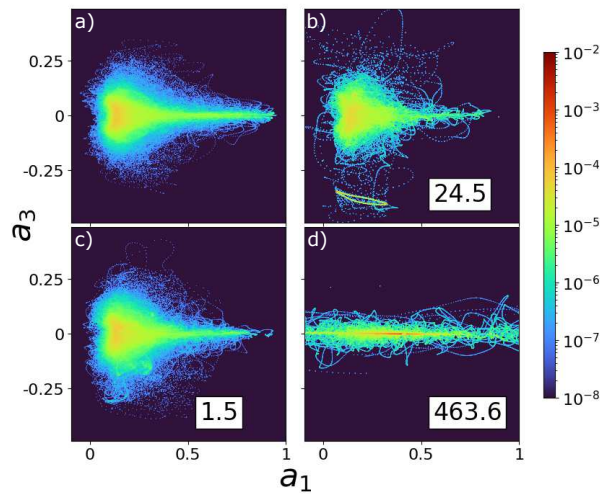}
\caption{(a) Joint probability density function (jPDF) of $a_1$ and $a_3$; (b-d), predictions of the MFE model by the single- and three-chart models and by the Koopman model, respectively, with the mean squared error (MSE) of the reconstruction, normalized by the MSE between two jPDFs produced by the MFE model with unique initial conditions, inset. Note the logarithmic scale.}
\label{jPDF_a3}
\end{figure}

Now we examine the ability of the data driven-model to correctly capture the structure of the extreme events.
An extreme event can be identified by a growth in the first MFE amplitude, which represents the mean shear, with a corresponding decrease in the remaining eight amplitudes, which capture the turbulent fluctuations.
In Fig.~\ref{jPDF_a3}.a, we show the joint probability density function (jPDF) of $a_1$ and $a_3$ for the reconstruction of the ensemble of trajectories from data set B, which is similar in distribution to that created by the turbulent portion of the training data.
The extreme events can be seen as the long tail extending to the right toward the laminar state $a_1=1, a_3=0$.
The prediction of the single-chart model, shown in \ref{jPDF_a3}.b fails to accurately capture the structure of the extreme events, with the tail almost entirely absent.
By contrast, the three-chart model, shown in \ref{jPDF_a3}.c, captures the structure of the extreme events well, accurately reproducing the shape of the joint probability density function.
The Koopman model, shown in \ref{jPDF_a3}.d, entirely fails to accurately capture the structure of the joint probability function. 
The accuracy of the reconstruction of the jPDFs is quantified by calculating the relative mean squared error (MSE) of the reconstructed jPDF, which is defined as the MSE between the jPDF of test data set and the reconstruction normalized by the MSE between two jPDFs produced by the MFE model from two sets of unique initial conditions.
The relative MSE of the reconstruction by the three-chart model is only 1.5, indicating excellent reconstruction of the jPDF.
The relative MSEs of the reconstructions by the single-chart model and the Koopman operator, however, are much larger at 24.5 and 463.6, respectively, reflecting the much poorer reconstructions.

We now examine the ability of the single- and multi-chart models to forecast an extreme event, defined by the kinetic energy of the time series increasing to $KE > 0.1$. 
To analyze the ability to predict quasi-laminarization events, each time series in data set A, which contained 90 extreme events, was segmented into time windows of duration $0.5\tau_L$ and analyzed for the presence of an extreme event (i.e., $KE$ exceeding $0.1$ in the window), where the window at $\tilde{t}$ refers to the time between $\tilde{t}$ and $\tilde{t}+0.5\tau_L$.
The exact solution and data-driven models were then compared to determine if each predicted whether an extreme event occurred.
If an extreme event occurred in both the exact solution and the model predictions, this was labeled as a \textit{true positive} ($TP$).
If the exact solution  exhibited an extreme event, but the data-driven model failed to forecast one, this was labeled as a \textit{false negative} ($FN$).
If the model predicted an extreme event when the exact solution showed none, it was identified as a \textit{false positive} ($FP$). \citep{Racca2022}
The total number of each identification type in each window was tabulated and the  \textit{F-score}, $F$, was calculated in each window, where
$F = (1+\frac{FP+FN}{2TP})^{-1}$.

\begin{figure}
\centering
\includegraphics[width=0.45\textwidth]{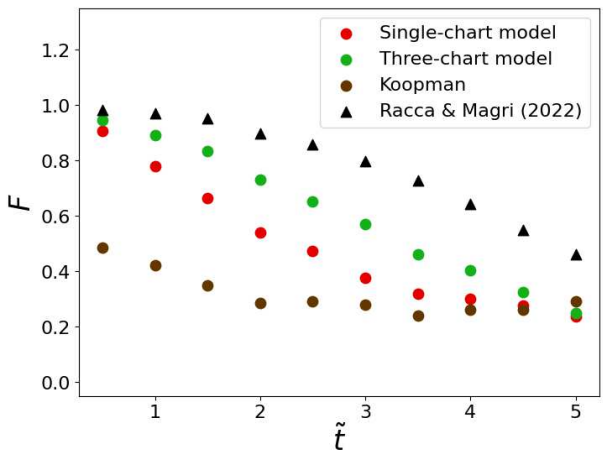}
\caption{F-score of extreme event forecasting of the MFE model by the single- and three-chart models and by the Koopman model, with comparison to prior study by \citet{Racca2022}.}
\label{F-score}
\end{figure}

Fig.~\ref{F-score} shows the F-score as a function of prediction time for the single- and multi-chart models, as well as a comparison to results from \citet{Racca2022} using an echo state network; this study did not perform quantitative trajectory comparisons, and as such a similar comparison can not be made for Fig.~\ref{Ensemble_Averaged_Error}.
Both the single- and multi-chart models produce more accurate forecasts of extreme event occurrences than the Koopman model, with the F-score of the Koopman model falling below 0.5 at all time windows.
The multi-chart model outperforms the single-chart model, more accurately forecasting extreme events at all prediction times.
Our multi-chart model compares favorably to the (non-Markovian) echo state network developed by \citet{Racca2022} at short times, while the accuracy falls below the improved predictive abilities of the ESN at longer prediction times.

Finally, we  determine the ability of the data-driven models to forecast the lifetime of the turbulence before permanent laminarization.
At long times, all time series generated by the MFE model at the given parameters collapse to the laminar fixed point; the lifetime of each time series is dependent on the initial condition, with the probability of remaining in the turbulent state approaching zero at long times.
At $Re \lesssim 320$,  the probability that a given time series remains in the turbulent state for a duration $t$, known as the survival function $S(t)$, takes the form \citep{Moehlis2004, Pershin2021} $S(t; Re) = \exp \left[ \frac{t-t_0}{\tau_S(Re)} \right]$, where $t_0$ is the time delay caused by the approach to the attractor and $1/\tau_S(Re)$ is the $Re$-dependent decay rate.
At $Re \gtrsim 320$, the distribution, particularly at long lifetimes, is known to deviate from an exponential decay, requiring increased time to laminarize.

\begin{figure}
\centering
\includegraphics[width=0.45\textwidth]{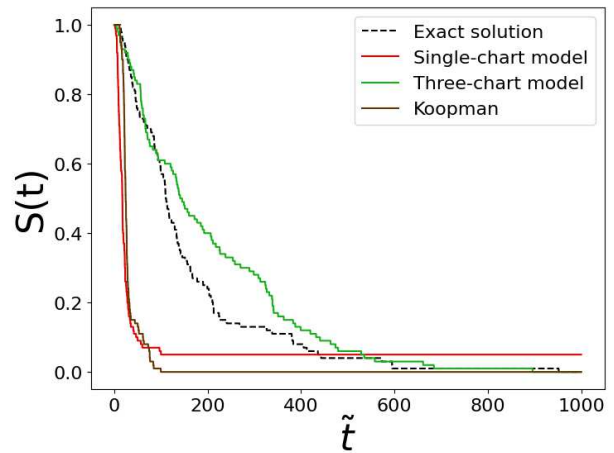}
\caption{Lifetime distribution of data and the reconstruction from the MFE model by the single- and three-chart models and by the Koopman model.}
\label{Lifetime_Distribution}
\end{figure}

Here, we define a laminarization event as a high-energy state ($KE>0.1$) for which the kinetic energy over 1 $\tau_L$ levels off.
The survival function, $S(t)$, is shown in Fig.~\ref{Lifetime_Distribution} for the test data set and the one- and three-chart models.
The test data set consists of 100 time series of varying lengths and has a mean lifetime of 169 $\tau_L$, within 3\% of the mean lifetime of the training data set (164 $\tau_L)$; this will henceforth be denoted as data set C.
The one-chart model and Koopman model produce poor predictions of the lifetime distribution, vastly underestimating the lifetimes of the turbulent state, with mean lifetimes of $19 \tau_L$ and $30 \tau_L$, respectively.
The three-chart model produces a much more accurate representation of the lifetime distribution.
The predicted distribution closely matches the exact solution for $\tilde{t}$ up to about $150$, while overestimating the lifetimes at longer times, and predicts an average lifetime of 200 $\tau_L$, overestimating the true result by less than 20\%.
It should be emphasized that we are measuring time here in units of Lyapunov time, so the inaccuracy of $S(t)$ in the three-chart model only arises at extremely long times.

\section{Conclusion}

In this paper, we have applied the CANDyMan \citep{Floryan2022} technique towards data-driven modeling of a dynamical system with extreme events: the MFE model \citep{Moehlis2004} for turbulent shear flow.
We have shown that clustering data sets and training multiple local data-driven models allows unique features of distinct data regimes (e.g.~extreme events) to be separately and more accurately captured by a multi-chart global model than in a conventional data-driven model.
Thus, multi-chart models were able to more accurately reproduce the evolution of this system, reducing forecasting error and improving reconstruction of the structure and frequency of extreme events.
Importantly, multi-chart models dramatically improved predictions of extreme event occurrences compared to the single-chart models used previously.
While we were not able to fully match the predicative capabilities of the non-Markvovian echo state networks in this regard, our Markovian models more accurately complies with the true nature of the underlying system being modeled.
Finally, we demonstrated the ability of multi-chart models to reconstruct the lifetime distribution of turbulent states, accurately predicting the distribution of survival times hundreds of Lyapunov times in the future.

In all cases, we have shown that our model outperforms a Koopman approach (EDMD-DL), in which all the observables are evolved linearly, in terms of short and long time tracking of the dynamics. While the approach captures key time-averaged quantities such as the mean velocity profile and Reynolds shear stress, its predictions of dynamics are less robust. The performance of the Koopman approach appears consistent with other studies (e.g.~\cite{Brunton2017}), where nonlinear modifications seem to be required to effectively model systems with chaotic dynamics.

An open question in the use of multi-chart models generated through the CANDyMan technique concerns the number of charts necessary for adequately capturing a dynamical system with extreme events.
In this study, we have observed that multi-chart data-driven models, specifically a three-chart model, can forecast one such system more accurately than a single-chart model.
Given the clustering of the training data into three charts, shown in Fig.~\ref{Clustering}, the desired number of charts should be sufficient to separate the extreme from the non-extreme states with a transition region between them.
This suggests that a dynamical system with one extreme state, such as the MFE model, would require at least three charts for optimal forecasting, whereas a dynamical system with multiple types of extreme events, such as weather systems, would necessitate more.
We have, however also observed that the improvement in predictive capabilities decreases above a certain number of charts, suggesting that the advantage gained through multi-chart models is limited and does that not increase indefinitely with additional charts.
In fact, the performance of our data-driven models decrease above three charts, suggesting that an unnecessarily large number of charts may hinder forecasting.
Observing the clustering shown in figure \ref{Clustering}, it can be seen that, between three and five clusters, the additional clustering does not further segment the extreme and non-extreme states, but rather separates only the non-extreme state into additional regions.
Further clustering of the data reduces the total data used to train any individual local data-driven model, potentially decreasing the performance of the local model.
In this case, the additional local models describe similar regions of data already well described by a single local model, and as such the decreased training data worsens the individual local models and thereby the global multi-chart model.
Accordingly, the optimal number of charts used to train a dynamical system with extreme events should allow clustering to separate out distinct features, while not further segmenting already distinct regions.
Further work is necessary to systematically understand how to best choose the number of a charts for a given data set.

Now that we have seen that  CANDyMan  has improved the performance of data-driven models forecasting a low-dimensional dynamical system with extreme events, future investigations should determine its applicability to higher-dimensional systems. 
As has been previously shown, the use of a charting technique such as CANDyMan allows improved dimension reduction through the use of autoencoder neural networks, capturing the intrinsic dimensionality of dynamical systems \citep{Floryan2022}.
For high-dimensional dynamical systems with intermittency, such as turbulent fluid flows, the application of  CANDyMan could not only aid in improved dimension reduction, but also produce more accurate forecasting than conventional single-chart techniques.

\section*{Acknowledgments}
\noindent This work was supported by ONR N00014-18-1-2865 (Vannevar Bush Faculty Fellowship). We gratefully acknowledge Daniel Floryan for helpful discussions.

\bibliography{apssamp}


\end{document}